\documentclass[%
 reprint,
 amsmath,amssymb,
 aps,
]{revtex4-2}

\usepackage{graphicx}
\usepackage{dcolumn}
\usepackage{bm}
\usepackage{ulem, color}

\newcommand\scalemath[2]{\scalebox{#1}{\mbox{\ensuremath{\displaystyle #2}}}}
\newcommand{\rev}[1]{\textcolor{black}{ #1}}

\begin{document}

\preprint{APS/123-QED}

\title{Entropy production of Multivariate Ornstein-Uhlenbeck processes correlates with consciousness levels in the human brain}


\author{Matthieu Gilson$^{1}$, Enzo Tagliazucchi$^{2}$, Rodrigo Cofré$^{3,4}$ }
\affiliation{$^{1}$Institut de Neurosciences des Systèmes INSERM-AMU, Marseille, France\\
$^{2}$Physics Department University of Buenos Aires and Buenos Aires Physics Institute Argentina\\
$^{3}$ CIMFAV-Ingemat, Facultad de Ingeniería, Universidad de Valparaíso, Chile\\
$^{4}$ Institute of Neuroscience (NeuroPSI-CNRS) Paris-Saclay University  Gif sur Yvette 91400, France 
}%


\begin{abstract}
Consciousness is supported by complex patterns of brain activity which are indicative of irreversible non-equilibrium dynamics. While the framework of stochastic thermodynamics has facilitated the understanding of physical systems of this kind, its application to infer the level of consciousness from empirical data remains elusive. We faced this challenge by calculating entropy production in a multivariate Ornstein-Uhlenbeck process fitted to fMRI brain activity recordings. To test this approach, we focused on the transition from  wakefulness to deep sleep, revealing a monotonous relationship between entropy production and the level of consciousness. Our results constitute robust signatures of consciousness while also advancing our understanding of the link between consciousness and complexity from the fundamental perspective of statistical physics. 
\end{abstract}

\maketitle


Animal cognition is the most sophisticated example of information processing found in biological and technological systems \citep{piccinini2011information}. Consciousness, understood as the capacity to sustain subjective experience, can be considered a property that emerges when a sufficiently high level of complex cognitive processing is achieved \citep{seth2022theories}. From the perspective of physics, consciousness and cognition seem unlikely to emerge from regular and predictable systems, such as those which are in  thermodynamic equilibrium and obey the detailed balance equations  \citep{Lynn2021}. Instead, recent research draws a close parallel between the level of consciousness and the entropy production rate of brain activity time series, highlighting temporal irreversibility as a landmark feature of conscious information processing \citep{SanzPerl2021, Fuente2021, munoz2020general}. These results suggest a close link between consciousness and non-equilibrium dynamics, prompting a rigorous evaluation from the perspective of stochastic thermodynamics. 

In spite of these exciting results, the direct estimation of entropy production from neural activity recordings is undermined by insufficient spatio-temporal sampling, leading to the adoption of heuristics and approximations which lack rigorous justification \citep{Lynn2021, SanzPerl2021}. To circumvent these limitations, we adopted a framework based on Multivariate Ornstein-Uhlenbeck (MOU) processes, that are widely used for modeling the multivariate dynamics of time series. 
The importance of MOU derives from the fact that it is the only continuous stationary stochastic process that is simultaneously Gaussian and Markovian. The MOU process is at the heart of many models used to fit fMRI data and to interpret them in terms of whole-brain communication \citep{Gilson2016,Friston2019,Frassle2021}, in line with the present methodology. We first characterize the non-equilibrium steady state of a generic MOU process. The irreversibility of the process is encoded in the antisymmetric part of the Onsager matrix, while the linearity of the Langevin equations allows us to derive closed-form expression for the entropy production rate in terms of the matrices that define the MOU. As a result, we  obtained a model-based estimation of the entropy production rate for the MOU fitted to fMRI data of subjects transitioning different levels of consciousness during the descent from wakefulness to deep sleep.

\ 

\section{Multivariate Ornstein-Uhlenbeck process}

We consider the MOU process closely following the notation in previous work \citep{Godreche2019}:
\begin{equation}\label{mou-vec}
\frac{\mathrm{d} \boldsymbol{x}(t)}{\mathrm{d} t}
=
-\boldsymbol{B} \, \boldsymbol{x}(t) 
+ \boldsymbol{\eta}(t) \ .
\end{equation}
Boldfaced symbols denote vectors and matrices. The inputs $\boldsymbol{\eta}(t)$ correspond to Gaussian white noise with covariance
\begin{equation}
\left\langle
\boldsymbol{\eta}(t) \, \boldsymbol{\eta}^{\mathrm{T}}\left(t^{\prime}\right)
\right\rangle_t
=
2\boldsymbol{D} \, \delta\left(t-t^{\prime}\right)
\ .
\end{equation}
The angular brackets indicate the mathematical expectation over time and the superscript $\mathrm{T}$ the transpose for vectors or matrices. The $N$-dimensional MOU process is thus defined by two real $N \times N$ matrices, the input covariance matrix $\boldsymbol{D}$, which is symmetric with positive eigenvalues, and the friction matrix $\boldsymbol{B}$, which is not symmetric in general.

\subsection{Description of the state evolution}

Knowing the initial condition $\boldsymbol{x}(0)$ and the realization of the stochastic input $\boldsymbol{\eta}$ over time, the trajectory of the solution of the Eq. \eqref{mou-vec} is given by:
\begin{equation}
\boldsymbol{x}(t)
=
\boldsymbol{G}(t) \, \boldsymbol{x}(0)
+ \int_{0}^{t} \boldsymbol{G}(t-s) \, \boldsymbol{\eta}(s) \, \mathrm{d} s \ ,
\end{equation}
where $\boldsymbol{G}(t)=\mathrm{e}^{-\boldsymbol{B} t}$ is the Green's function, also known as propagator. 
In addition to its mean value $\langle\boldsymbol{x}(t)\rangle = \boldsymbol{G}(t) \, \boldsymbol{x}(0)$, the process is also characterized by its  covariance matrix $\boldsymbol{S}(t,t') = \langle \boldsymbol{x}(t) \boldsymbol{x}^\mathrm{T}(t^{\prime}) \rangle$.
The zero-lag covariance, denoted by $\boldsymbol{S}(t,t)$, obeys the following deterministic differential equation:
\begin{equation} \label{evol-cov-eq}
\frac{\mathrm{d} \boldsymbol{S}(t,t)}{\mathrm{d} t}
=
- \boldsymbol{B} \, \boldsymbol{S}(t,t)
- \boldsymbol{S}(t,t) \, \boldsymbol{B}^{\mathrm{T}}
+ 2 \boldsymbol{D} \ .
\end{equation}
Meanwhile, the lagged covariance with $t^{\prime}>t$ exhibits an exponential decay as a function of the lag $t^{\prime} - t$:
\begin{equation} \label{lag-cov-eq}
\boldsymbol{S}(t,t^{\prime})
=
\boldsymbol{S}(t, t) \, e^{-\boldsymbol{B}^\mathrm{T}(t^{\prime}-t)} \ . 
\end{equation}



A standard method for analysing Eq.\eqref{mou-vec}, consists in describing the evolution of the probability distribution $P(\boldsymbol{x}, t)$ via the Fokker–Planck equation:
\begin{equation}\label{fpe}
\frac{\partial P(\boldsymbol{x}, t)}{\partial t} 
=
\nabla \cdot\left[
\boldsymbol{B} \, \boldsymbol{x}(t) \, P(\boldsymbol{x}, t)
+
\boldsymbol{D} \, \nabla P(\boldsymbol{x}, t)
\right] \ ,
\end{equation}
where $\nabla$ denotes the spatial derivative with respect to $\boldsymbol{x}$.
Eq.~\eqref{fpe} can be rewritten as a continuity equation of the form
\begin{equation}
\frac{\partial P(\boldsymbol{x}, t) }{ \partial t } 
+
\nabla \cdot \boldsymbol{J}(\boldsymbol{x}, t)
=
0 \ .
\end{equation}
with the following expression for the probability current (or flux)
\begin{equation} \label{eqj}
\boldsymbol{J}(\boldsymbol{x}, t)
=
-\boldsymbol{D} \nabla P(\boldsymbol{x}, t)-\boldsymbol{B}\boldsymbol{x}(t) P(\boldsymbol{x}, t)
\end{equation}

\subsection{Stationary state and probability current}

The Gauss-Markov property of the Ornstein-Uhlenbeck process ensures that the mean and covariances converge exponentially fast toward their respective fixed points, provided the eigenvalues of $\boldsymbol{B}$ (which may be complex) have positive real part. The stationary state of the MOU process exhibits Gaussian fluctuations around a mean equal to zero. This corresponds to the time-independent multivariate probability density
\begin{equation} \label{stat-sol}
P(\boldsymbol{x})
=
\frac{1}{(2 \pi)^{N / 2} (\operatorname{det} \boldsymbol{S})^{1 / 2}} \exp \left(-\frac{1}{2} \boldsymbol{x}^{\mathrm{T}} \boldsymbol{S}^{-1} \boldsymbol{x}\right) \ ,
\end{equation}
where $\boldsymbol{S}$ denotes the fixed point of the zero-lag covariance matrix $\boldsymbol{S}(t,t)$.
From Eq.~\eqref{stat-sol}, the gradient of $P(\boldsymbol{x})$ simply reads
\begin{equation}
\nabla P(\boldsymbol{x}) 
=
\frac{\partial  P(\boldsymbol{x})}{\partial \boldsymbol{x}}
=
- P(\boldsymbol{x}) \, \boldsymbol{S}^{-1} \, \boldsymbol{x}
\end{equation}
From Eq.~\eqref{eqj}, the stationary probability current $\boldsymbol{J}(\boldsymbol{x})$ can thus be rewritten in a compact form 
\begin{eqnarray}\label{jmuxp}
\boldsymbol{J}(\boldsymbol{x})
& = &
\boldsymbol{D} \, P(\boldsymbol{x}) \, \boldsymbol{S}^{-1} \, \boldsymbol{x}
- \boldsymbol{B} \, \boldsymbol{x} \, P(\boldsymbol{x})
\\
& = &
\boldsymbol{\mu} \, \boldsymbol{x} \, P(\boldsymbol{x})
\nonumber \ ,
\end{eqnarray}
with
\begin{equation} \label{prob-curr}
\boldsymbol{\mu}= \boldsymbol{D} \, \boldsymbol{S}^{-1}-\boldsymbol{B}
\end{equation}

\subsection{Entropy production rate}

Going a step further, the (ir)reversibility can be described using thermodynamic variables evaluated for the dynamic process.
Using the well-known definition for entropy for the probability distribution $P(\boldsymbol{x},t)$, now considering its time dependent version, we have

\begin{equation}
e[P]=-\int_{\mathbb{R}^{n}} P(\boldsymbol{x}, t) \, \log P(\boldsymbol{x}, t) \, \mathrm{d} \boldsymbol{x}\ .
\end{equation}

It can be shown that the rate of the increase of entropy over time can be decomposed into two factors, namely $\dot{e}[P]= EPR -HDR$, where EPR is the entropy production rate and HDR the heat-dissipation rate \citep{Qian2001,  Godreche2019, Cofre2018}. The EPR is the main quantity of interest here, which we denote by $\Phi$.
Now calculating $\Phi$ for the time-independent distribution $P(\boldsymbol{x})$, we have

\begin{equation}\label{pitd}
\Phi= \int \frac{\boldsymbol{J}^{\mathrm{T}}(\boldsymbol{x})  \boldsymbol{D}^{-1} \boldsymbol{J}(\boldsymbol{x})}{P(\boldsymbol{x})} \mathrm{d} \boldsymbol{x} = \left\langle \boldsymbol{\Pi}^{\mathrm{T}} \boldsymbol{D} \boldsymbol{\Pi}\right\rangle
\end{equation}

where $\boldsymbol{\Pi}$ is called the the thermodynamic force and is related to $\boldsymbol{J}$  by the Onsager’s reciprocal relations \citep{Qian2001}: 
\begin{equation}\label{pi-j}
\boldsymbol{\Pi}=\frac{\boldsymbol{D}^{-1} \boldsymbol{J}}{P}
\end{equation}

The heat-dissipation rate can be computed as follows:
\begin{equation}
HDR=\int_{\mathbb{R}^{n}} \boldsymbol{D}^{-1} \boldsymbol{B}\boldsymbol{x}  \cdot \boldsymbol{J} \mathrm{d} x
\end{equation}

In the context of the stationary MOU diffusion processes, a general expression for the entropy production rate per unit time in the stationary state is the following \citep{Qian2001, Qian2002, Godreche2019}

\begin{equation}\label{eqphi}
\scalemath{0.82}{\Phi= \int\left(\nabla \log P(\boldsymbol{x})-\boldsymbol{D} \boldsymbol{B}\boldsymbol{x}\right)^{T} \boldsymbol{D}\left(\nabla \log P(\boldsymbol{x})-\boldsymbol{D}^{-1} \boldsymbol{B}\boldsymbol{x}\right) P(\boldsymbol{x}) d \boldsymbol{x}}
\end{equation}

\noindent
which can be obtained from (\ref{prob-curr}), (\ref{pitd}) and (\ref{pi-j}) as follows:

\begin{equation}\label{eq1}
\begin{aligned}
\mu & = \boldsymbol{D}\boldsymbol{S}^{-1}-\boldsymbol{B} \\
\boldsymbol{D}^{-1}\mu & = \boldsymbol{S}^{-1}-\boldsymbol{D}^{-1}\boldsymbol{B} && \boldsymbol{D}^{-1}\cdot \\
\boldsymbol{D}^{-1}\mu \boldsymbol{x} & = (\boldsymbol{S}^{-1}-\boldsymbol{D}^{-1}\boldsymbol{B}) \boldsymbol{x}  && \cdot \boldsymbol{x}\\
\boldsymbol{D}^{-1}\mu \boldsymbol{x}P & = (\boldsymbol{S}^{-1}-\boldsymbol{D}^{-1}\boldsymbol{B}) \boldsymbol{x}P && \cdot P \\
\boldsymbol{D}^{-1}  \boldsymbol{J}& = (\boldsymbol{S}^{-1}-\boldsymbol{D}^{-1}\boldsymbol{B}) \boldsymbol{x}P &&  \text{from } (\ref{jmuxp})\\
\boldsymbol{\Pi}& = (\boldsymbol{S}^{-1}-\boldsymbol{D}^{-1}\boldsymbol{B}) \boldsymbol{x} && \text{from } (\ref{pi-j})\\
\boldsymbol{\Pi}& = \boldsymbol{S}^{-1}\boldsymbol{x}-\boldsymbol{D}^{-1}\boldsymbol{B} \boldsymbol{x}
\end{aligned}
\end{equation}

Now, as $\nabla \log P(\boldsymbol{x})=\boldsymbol{S}^{-1}\boldsymbol{x}$, we obtain (\ref{eqphi}). From (\ref{pitd})

$$
\left\langle \boldsymbol{\Pi}^{\mathrm{T}} \boldsymbol{D} \boldsymbol{\Pi}\right\rangle = \left\langle\boldsymbol{x}^{\mathrm{T}}\left(\boldsymbol{D}^{-1} \boldsymbol{B}-\boldsymbol{S}^{-1}\right)^{\mathrm{T}}  \boldsymbol{D}\left(\boldsymbol{D}^{-1} \boldsymbol{B}-\boldsymbol{S}^{-1}\right) \boldsymbol{x}\right\rangle,
$$

\noindent
we obtain that 

\begin{equation}\label{phi}
\Phi=\left\langle\boldsymbol{x}^{\mathrm{T}}\left(\boldsymbol{D}^{-1} \boldsymbol{B}-\boldsymbol{S}^{-1}\right)^{\mathrm{T}}  \boldsymbol{D}\left(\boldsymbol{D}^{-1} \boldsymbol{B}-\boldsymbol{S}^{-1}\right) \boldsymbol{x}\right\rangle,
\end{equation}
\noindent
where the average is taken over the stationary state of the process. From this equation we can verify that when $\boldsymbol{S}=\boldsymbol{B}^{-1}  \boldsymbol{D}$, then $\Phi =0$. 

Following previous results \citep{Risken1996,Qian2001}, a sufficient condition for the MOU process in Eq.~\eqref{mou-vec} to be a time reversible stationary process corresponds to a specific relation between the matrices $\boldsymbol{B}$ and $\boldsymbol{D}$: 
\begin{equation} \label{symm-con}
\boldsymbol{B} \, \boldsymbol{D}
= 
\boldsymbol{D} \, \boldsymbol{B}^{\mathrm{T}}.
\end{equation}
To quantify the time (ir)reversibility of the MOU process, it is advantageous to examine the Onsager matrix $\boldsymbol{L}$ reparameterized using the matrices $\boldsymbol{B}$,  $\boldsymbol{D}$, and the  pairwise zero-lag covariance $\boldsymbol{S} = \langle \boldsymbol{x}(t) \boldsymbol{x}^\mathrm{T}(t) \rangle_t$:
\begin{eqnarray} \label{symm-antisymm}
\boldsymbol{L}
& = & 
\boldsymbol{B} \, \boldsymbol{S}
= \boldsymbol{D} + \boldsymbol{Q} \ , 
\\
\boldsymbol{L}^{\mathrm{T}}
& = &
\boldsymbol{S} \, \boldsymbol{B}^{\mathrm{T}}
= \boldsymbol{D} - \boldsymbol{Q} 
\nonumber \ .
\end{eqnarray}
Here the antisymmetric part $\boldsymbol{Q}$ of $\boldsymbol{L}$ provides a measure for the irreversibility of the process. When the process is time reversible $\boldsymbol{Q}=0$ and $\boldsymbol{L}$ is symmetric. The following expression for the entropy production rate $\Phi$ can then be derived \rev{from the differential entropy of a multivariate Gaussian, which is a well defined quantity.} 

From equations \eqref{symm-antisymm} and \eqref{prob-curr}, we have $\boldsymbol{D}^{-1} \boldsymbol{B}-\boldsymbol{S}^{-1}= \boldsymbol{D}^{-1} \boldsymbol{Q} \boldsymbol{S}^{-1}=- \boldsymbol{D}^{-1} \boldsymbol{\mu}$. Thus, from Eq.~\eqref{phi} considering that $\boldsymbol{S}$ and $\boldsymbol{D}$ are symmetric and $\boldsymbol{Q}$ is anti-symmetric we obtain:
\begin{equation}
\Phi=-\left\langle\boldsymbol{x}^{\mathrm{T}} \boldsymbol{S}^{-1} \boldsymbol{Q}  \boldsymbol{D}^{-1} \boldsymbol{Q} \boldsymbol{S}^{-1} \boldsymbol{x}\right\rangle=\left\langle\boldsymbol{x}^{\mathrm{T}} \boldsymbol{\mu}^{\mathrm{T}}  \boldsymbol{D}^{-1} \boldsymbol{\mu} \boldsymbol{x}\right\rangle
\end{equation}

The entropy production rate $\Phi$ is non-negative. It is strictly positive if the process is irreversible, and it vanishes only if the process is reversible. Since the stationary state of the MOU is Gaussian with covariance matrix $\boldsymbol{S}$, we have the following property: $\left\langle\boldsymbol{x}^{\mathrm{T}} \boldsymbol{A} \boldsymbol{x}\right\rangle=\operatorname{tr}(\boldsymbol{S} \boldsymbol{A})$, and so
\begin{equation}\label{Eprod-a}
\Phi =-\operatorname{tr}\left(\boldsymbol{S}^{-1} \boldsymbol{Q}  \boldsymbol{D}^{-1} \boldsymbol{Q}\right) =\operatorname{tr}\left(\boldsymbol{S} \boldsymbol{\mu}^{\mathrm{T}}  \boldsymbol{D}^{-1} \boldsymbol{\mu}\right),
\end{equation}
which can be written into the following equivalent expressions, that does not involve the covariance matrix $\boldsymbol{S}$ nor its inverse explicitly:
\begin{equation}\label{Eprod}
\Phi=\operatorname{tr}\left(\boldsymbol{B}^{\mathrm{T}}  \boldsymbol{D}^{-1} \boldsymbol{Q}\right)=-\operatorname{tr}\left(\boldsymbol{D}^{-1} \boldsymbol{B} \boldsymbol{Q}\right),
\end{equation}

The entropy production rate $\Phi$ provides a scalar measure for the (ir)reversibility of the whole network process, vanishing only if the process is reversible.



\section{Methods}

\subsection{Empirical covariance from fMRI data}

The model is fitted to reproduce the two covariance matrices calculated from the empirical BOLD signals, with zero lag and a lag equal to 1~TR: 
\begin{eqnarray} \label{emp-cov}
    \widehat{S}_{i j}(0)
    & = &
    \frac{1}{T-2} \sum_{1 \leq t \leq T-1}\left[x_{i}(t)-\bar{x}_{i}\right]
    \left[x_{j}(t)-\bar{x}_{j}\right] \ , 
    \\
    \widehat{S}_{i j}(1)
    & = &
    \frac{1}{T-2} \sum_{1 \leq t \leq T-1} \left[x_{i}(t)-\bar{x}_{i}\right]
    \left[x_{j}(t+1)-\bar{x}_{j}\right] \ .
\end{eqnarray}
Here $\bar{x}_{i}$ denotes the mean empirical signal: $\bar{x}_{i}=\frac{1}{T} \sum_{t} x_{i}(t)$ for all $i$, which is used to center the data as all variables $x_{i}$ have mean zero in the model. These are the empirical counterparts of the model covariances $S_{ij}(t,t)$ and $S_{ij}(t,t+1)$ averaged over time $t$.

\subsection{Parameter estimation of the MOU process}

We fit the MOU process from the fMRI time series data for each subject in each sleep condition. We rely on a recent estimation method that tunes the MOU model such that its covariance structure reproduces the matrices in Eq.~\eqref{emp-cov}, optimizing its parameters the Jacobian matrix $-\boldsymbol{B}$ as well as the input covariance matrix $2 \boldsymbol{D}$ \citep{Gilson2016}. Importantly, this optimization procedure incorporates topological constraints on $\boldsymbol{B}$, adjusting only existing anatomical connections, also keeping the input cross-covariances $D_{ij} = 0$ for $i \neq j$. Note that our current notation corresponds to a previous publication \citep{Gilson2016}, using the following $-\boldsymbol{B} \leftrightarrow \boldsymbol{J}$ and $2\textbf{D} \leftrightarrow \Sigma$; note that $-\boldsymbol{B} \leftrightarrow \boldsymbol{J}^{\mathrm{T}}$ in the subsequent paper \citep{Gilson2020}.

The model is first calibrated by calculating the time constant $\tau$ from the empirical signals. 
\begin{equation}
\tau
=
-\frac{N}{\sum_{1 \leq i \leq N} a\left(v_{i} \mid u\right)} \ ,
\end{equation}
where $a\left(v_{i} \mid u\right)$ is the slope of the linear regression of $v_{i}=\left[\log \left(\widehat{S}_{i i}^{0}\right), \log \left(\widehat{S}_{i i}^{1}\right)\right]$ by $u=[0,1]$.

We rely on a gradient descent to iteratively adjust $\boldsymbol{B}$ and $\boldsymbol{D}$ until reaching the best fit \citep{Gilson2016}.
At each optimization step, we calculate the model counterparts of the covariance matrices in Eq.~\eqref{emp-cov} $\boldsymbol{S}(0)$ and $\boldsymbol{S}(1)$, assuming stationarity over each fMRI session. They can be calculated by solving the Lyapunov equation using e.g.\ the Bartels-Stewart algorithm, which yields here
\begin{equation} \label{eq-sylv}
\boldsymbol{B} \, \boldsymbol{S}(0) + \boldsymbol{S}(0) \, \boldsymbol{B}^{\mathrm{T}}
=
2 \boldsymbol{D} \ ,
\end{equation}
once again equating the derivative with zero in Eq.~\eqref{evol-cov-eq}, and the equation involving the propagator. We calculate the lagged covariance rewriting Eq.~\eqref{lag-cov-eq} for the time-lag equation here as
\begin{equation} \label{cov1-eq}
\boldsymbol{S}(1)
=
\boldsymbol{S}(0) \, e^{\boldsymbol{-B}^\mathrm{T}} \ .
\end{equation}
We then calculate the difference between the model and empirical covariances, $\Delta \boldsymbol{S}(t) = \hat{\boldsymbol{S}}(t) - \boldsymbol{S}(t)$ with $t \in \{0, 1\}$.
The parameter update is given by differentiating Eqs.~\eqref{cov1-eq} and \eqref{eq-sylv}:
\begin{eqnarray} \label{update_B_D}
\Delta \boldsymbol{B} 
& = &
\epsilon_B \left[ \boldsymbol{S}(0) \right]^{-1} \left[ \Delta \boldsymbol{S}(0) - \Delta \boldsymbol{S}(1) \, e^{\boldsymbol{B}^\mathrm{T}} \right]
\ , \\
\Delta \boldsymbol{D}
& = &
\epsilon_D \boldsymbol{B} \, \Delta \boldsymbol{S}(0) 
+ \epsilon_D \Delta \boldsymbol{S}(0) \, \boldsymbol{B}^{\mathrm{T}} 
\nonumber \ ,
\end{eqnarray}
with $\epsilon_B$ and $\epsilon_D$ small learning rates.
The best fit corresponds to minimising the squared norm of both $\Delta \boldsymbol{S}(0)$ and $\Delta \boldsymbol{S}(1)$.


\subsection{MOU-based anatomo-functional model to fit empirical fMRI data}
\begin{figure}[t!]
\centering
\includegraphics[width=1\columnwidth]{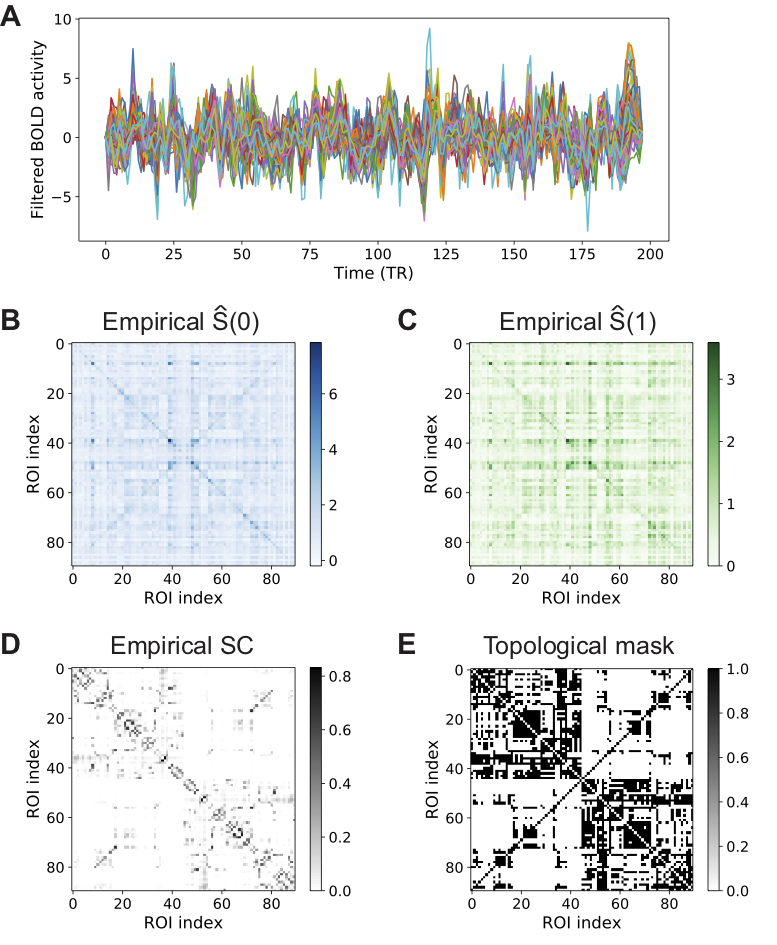}
\caption{ A) Example of the filtered BOLD time series with $198$ \rev{repetition times (TR) of 2 seconds}, corresponding to the 90 ROIs of the AAL parcellation during wakefulness of one participant. 
B-C) Functional connectivity matrices calculated from the filtered BOLD signals in panel A, $\hat{S}(0)$ with zero lag and $\hat{S}(1)$ with a lag of one timestep (TR=2~s). These matrices are used in the objective functions used to fit the anatomo-functional model.
D) Generic structural connectivity (SC) obtained from DTI data as described in \citep{Ipina2020}.
E) Mask for existing directional connections to constrain the topology of the $B$ matrix in the network model (symmetric here).
}
\label{fig1_emp_data}
\end{figure}

We fitted a MOU process to the time series of  blood oxygen level-dependent (BOLD) activity measured using fMRI for a whole-brain parcellation consisting of $N=90$ regions of interest (ROIs). The BOLD signals were recorded from 15 healthy participants during wakefulness and three sleep stages of progressively deeper unconsciousness (N1, N2, N3). Further details about the data preprocessing like detrending and filtering can be found in \citep{Tagliazucchi2014}. Example BOLD time series are illustrated in Fig.~\ref{fig1_emp_data}A. 
Fig.~\ref{fig1_emp_data}B-C show two functional connectivity matrices, here calculated as covariances with zero lag $\hat{\textbf{S}}(0)$ and lag of 1 timestep $\hat{\textbf{S}}(1)$. 
These matrices are the empirical counterparts of the model pairwise covariance $\textbf{S}(l)= \langle \boldsymbol{x}(t) \boldsymbol{x}^\mathrm{T}(t+l) \rangle_t$ with lag $l$, which is symmetric for $l=0$ and was denoted above by $\textbf{S}=\textbf{S}(0)$.

In this application, the activity $x_{i}$ of the MOU process describes the BOLD activity of node $i$. Its friction matrix $\textbf{B}$ quantifies the propagation of BOLD activity between ROIs, ignoring hemodynamics \rev{\cite{Gilson2016mb}}. Specifically, the diagonal elements $B_{ii}$ are related to a time constant $\tau$ (identical for all ROIs) and the off-diagonal elements $C_{ij} = -B_{ij}$ correspond to the concept of effective connectivity from ROI $j$ to ROI $i$ (excitatory when $C_{ij} > 0$):
\begin{equation}
-B_{ij}
=
- \frac{\delta_{ij}}{\tau}
+ C_{ij} 
\ ,
\end{equation}
where $\delta_{i j}$ is the Kronecker delta. The variance $D_{ii}$ reflects the fluctuation amplitude of ROI $i$. 

For each subject and condition, the model was fitted to reproduce the two covariance matrices calculated from the empirical BOLD signals $\hat{\textbf{S}}(0)$ and $\hat{\textbf{S}}(1)$ (see Fig.~\ref{fig1_emp_data}B-C). We used a recent estimation method based on gradient descent to iteratively adjust $\boldsymbol{B}$ and $\boldsymbol{D}$ until reaching the best fit \citep{Gilson2016}. At each optimization step, we calculate the model counterparts of the covariance matrices $\boldsymbol{S}(0)$ and $\boldsymbol{S}(1)$, assuming stationarity over each fMRI session. Importantly, this optimization procedure incorporates topological constraints on $\boldsymbol{B}$, adjusting only existing anatomical connections (see Fig.~\ref{fig1_emp_data}D-E), also keeping the input cross-covariances $D_{ij} = 0$ for $i \neq j$. Model fit is quantified by two measures: model error, defined using the matrix distance and Pearson correlation
between vectorized FC matrices (model versus data). All sleep states have Pearson correlation above 0.6, corresponding to an R2 of 0.36 (See fig S2 in the supplemetal material).

\begin{figure}[t!]
\centering
\includegraphics[width=1\columnwidth]{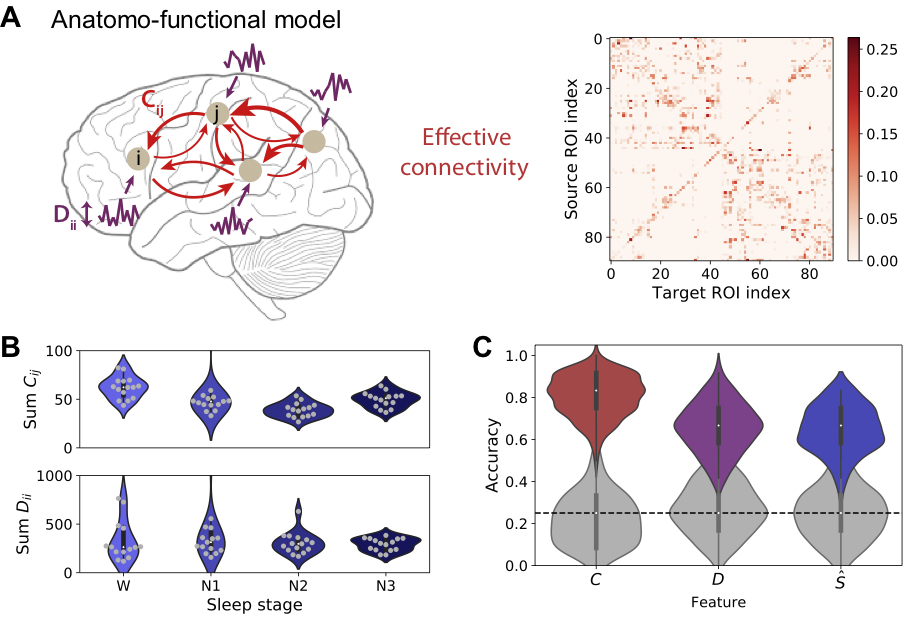}
\caption{
A) Our dynamic network model has two sets of optimized parameters: the matrix $\textbf{C}$ (effective connectivity), which describes the causal interaction between brain regions, and the input variance $\textbf{D}$, which represents the spontaneous activity of each brain region. Note that the topology of the matrix $\textbf{C}$ corresponds to the mask inferred from the SC data in Fig.~\ref{fig1_emp_data}D-E, but the weights are estimated from the empirical \rev{functional connectivity} (FC) matrices Fig.~\ref{fig1_emp_data}B-C, resulting in an anatomo-functional model.
B) Changes in total $\textbf{C}$ and $\textbf{D}$ weights across sleep stages (x-axis), pooled over the 15 subjects. The sleep stages are represented by the blue contrasts, from light for wake (W) to dark for the deepest sleep (N3).
C) Classification accuracy based on the model estimates, $\textbf{C}$ and $\textbf{D}$, and the empirical covariance matrices. The classifier is the multinomial logistic regression (MLR), which captures changes in individual features across sleep stages. The gray violin plots correspond to the chance-level accuracy calculated empirically by shuffling the labels of the sleep stages. }
\label{fig2_model}
\end{figure}

\begin{figure}[t!]
\centering
\includegraphics[width=0.5\textwidth]{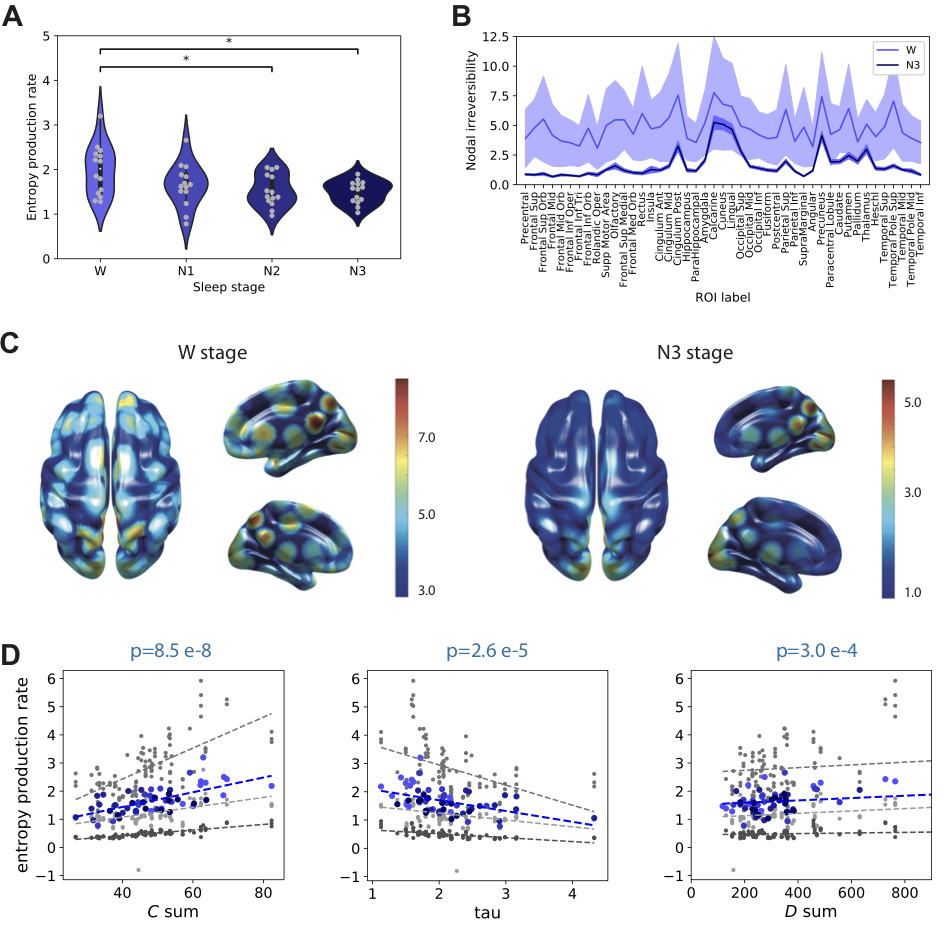}
\caption{
A) Violin plots comparing the entropy production rate across sleep stages. Same color coding used in previous plots. The average entropy production values across subjects are for the four sleep stages are $1.99$, $1.65$, $1.54$, and $1.49$, respectively. The stars indicate statistical significance for the Mann-Whitney test with $p<0.05$. B) Comparison of the nodal irreversibility for each ROI  (x-axis) between the W and N3 states (in light and dark blue, respectively). The plotted values correspond to the absolute value of sums over rows of $Q$, averaged for homotopic regions; error bars indicate the variability across subjects measured as the standard error of the mean. 
C) Heatmap plots of the nodal irreversibility on the cortical surface for the W and N3 sleep stages. Note the different color scales for the two stages, for the purpose of better readability.
D) \rev{To gain insight into the effects of the matrices $\textbf{C}$ and   $\textbf{D}$ on $\Phi$, we shuffle their values as a way to destroy their detailed structures (redistributing their values keeping the topology). We plot $\Phi$ for the estimated $\textbf{C}$ and $\textbf{D}$ matrices across subjects and sleep stages in blues (same color code used in previous plots), and the shuffled $\textbf{C}$ matrices (light gray), $\textbf{D}$ matrices (middle gray) and both (dark gray), as a function of the sum of $\textbf{C}$ values (left panel), the time constant $\tau$ (middle panel) and the sum of $\textbf{D}$ variances (right panel). }}
\label{fig3_epr}
\end{figure}

\subsection{Robust decoding of sleep stages from MOU parameters}
Following previous work \citep{Pallares2018, Gilson2020}, we used the scikit-learn Python library for the implementations of multinomial logistic regression (MLR) classifier. 
The input features corresponded to the vectorized $\textbf{C/D/S}$  matrices after discarding zero or redundant elements. We implemented a stratified cross-validation scheme with 80\% of the samples for the train set and 20\% for the test set, where the ratio of classes is the same in both sets. We also use the subject identity as ``group information'' to avoid mixing subject data between the train and test sets. In practice, we use 100 random splits of the data and report the distribution of the accuracies of the 100 splits. 

As illustrated in Fig.~\ref{fig2_model}B, both the empirical BOLD variances and the model estimates exhibit  global differences across the four sleep stages, although they do not exhibit a clear trend. These differences in global measures, which are averages over all ROIs, may hide more specific changes at the ROI level, as well as interactions between them. Supplementary Figure~S1 shows the good fit of the anatomo-functional model to fMRI data obtained for all sleep stages, with mean correlation between simulated and empirical FC matrices exceeding 0.6 for all stages. 
Fig.~\ref{fig2_model}C shows that the model estimates give good classification accuracy, both for $C$ (in red) and $D$ (in purple). This indicates that the model captures the differences in brain dynamics across the sleep stages. Notably, the matrix $C$ gives a better classification accuracy than the empirical functional connectivity $\hat{S}(0)$ (in blue), meaning that the model inversion is robust and captures refined information about the sleep stages. 
Note that the MLR has better accuracy than the 1-nearest-neighbor (1NN) in Suppl Fig~S1A, indicating that the changes across sleep stages concern specific features, i.e.\ connectivity weights ($\textbf{C}$) or nodal spontaneous activity ($\textbf{D}$), rather than their global profile.

\subsection{Reduced entropy production in the transition from wakefulness to deep sleep}
Using the condition-specific estimated parameters, we calculated the entropy production rate in the MOU model using Eq. \eqref{Eprod}. These results in Fig~\ref{fig3_epr}A show that entropy production decreases as a function of sleep depth, which in turn implies that dynamics become closer to equilibrium. 

The model-based approach allows us to dissect this phenomenon.
For all ROIs, we observe that the contribution to $\Phi$, as measured via the nodal irreversibility, defined as $\sum_j | Q_{ij} |$ for each ROI $i$, decreases, as illustrated in Fig~\ref{fig3_epr}B. This suggests that the reduction of $\Phi$ from W to N3 is a rather global phenomenon, but with a differentiated magnitude across brain regions. Notably, regions in the occipital lobes (cuneus, calcarine, lingual), as well as regions associated to hubs in the default-mode network (precuneus, post cingulate), and the thalamus, remain at a high level of nodal irreversibility in the deep sleep N3; these regions have been shown to exhibit sleep-related changes in previous studies \citep{Horovitz2009, Dang-Vu2010, Mirandola2013}. See Suppl Fig~S3 for a more detailed comparison across sleep stages.

Last, we examine how the model parameters $\textbf{C}$ and $\textbf{D}$ contribute to $\Phi$ and its reduction across sleep stages. Fig~\ref{fig3_epr}D shows a positive relationship between $\Phi$ and the sum of weights in $\textbf{C}$, as well as the sum of variances in $\textbf{D}$; conversely, a larger $\tau$ (directly calculated from the empirical BOLD signals) corresponds to a lower $\Phi$. Then we assess the importance of the detailed structures in the $\textbf{C}$ and $\textbf{D}$ estimates by randomizing them spatially, namely redistributing the total weight/variances across non-zero elements while keeping the same topology and overall sum. We observe the same trends with respect to the $\textbf{C}$ and $\textbf{D}$ sums, but shifted up or down depending on the surrogates in Fig~\ref{fig3_epr}C: randomizing $\textbf{C}$ (light gray) decreases slightly $\Phi$, whereas randomizing $\textbf{D}$ (middle gray) increases $\Phi$; randomizing both (dark gray) decreases $\Phi$. This indicates that $\Phi$ strongly depends on the detailed structures of the $\textbf{C}$ and $\textbf{D}$ estimates, being larger in the data than in the randomized surrogates. The opposing effects in randomizing $\textbf{C}$ and $\textbf{D}$ also suggest a balance implemented by the detailed brain dynamics, which results in a controlled level of $\Phi$. Together, our results hint at a positive relationship between the measured $\Phi$ and the different levels of consciousness.

\section{Discussion}
We measured the entropy production using our anatomo-functional MOU process associated to resting-state fMRI activity recorded from human subjects in different sleep stages. The advantage of our model-based approach is that the entropy production has a closed-form expression from first principles of stochastic thermodynamics for the MOU process, which is numerically fitted to the fMRI data. Our results show high entropy production rate in conscious wakefulness, i.e.\ correlating positively with the presumed level of cognitive processing. This is consistent with converging theoretical accounts that identify consciousness with an emergent property of a highly complex physical system \citep{seth2022theories}. These results are also consistent with previous findings relating entropy production with states of consciousness \citep{Fuente2021, SanzPerl2021, munoz2020general}, with the advantage that do not depend on heuristic approximations. Importantly, our approach allows for identifying the brain regions that contribute most to entropy production.
The fulfillment of detailed balance in the brain is scale-dependent \citep{Lynn2021}. At the large scale, its violation might relate to the large-scale circuit operations critical for healthy cognition and for the global broadcasting of information which is identified with the computational aspect of consciousness \cite{Dehaene1998}. Because of this, metrics related to the departure from detailed balance (such as entropy production rate) might offer valuable tools to determine levels of consciousness in brain-injured patients and other neurological populations.
In summary, assessing temporal irreversibility through entropy production of MOU processes derived from fMRI signals has the potential to highlight different states of consciousness and cognition. More generally, can bridge brain dynamics and thermodynamics, and ultimately help to understand fundamental questions about the brain and consciousness.

\textbf{Acknowledgements.} M.G. and R.C. would like to thank the organizers of the 2021 Spring School of the European Institute of Theoretical Neuroscience, the place where this project began to develop. MG was supported by the European Union’s Horizon 2020 Framework Programme for Research and Innovation under the Specific Grant Agreement No. 945539 (Human Brain Project SGA3), as well as
the Excellence Initiative of the German federal and state governments (ERS PF-JARA-SDS005). E.T. is supported by grants PICT2018-03103 and PICT-2019-02294 funded by Agencia I+D+I (Argentina), by a Mercator fellowship granted by the German Research Foundation and FONDECYT regular 1220995. R.C was supported by the Human Brain Project, H2020-945539. 

\section*{Supplemental material}
In this supplemental material, we present details on sleep stage decoding sleep stages, supplementary analysis, and a detailed description of the fMRI data.

\subsection{Decoding of sleep stages}

We use the same approach as in previous work \citep{Pallares2018, Gilson2020}. We rely on two usual classifiers: multinomial logistic regression (MLR) and the 1-nearest-neighbor (1NN). The features correspond to vectorized $\textbf{C/D/S}$ matrices after discarding zero or redundant elements. The MLR is a canonical tool for high-dimensional linear classification, which tunes a weight for each feature, thus selecting the important ones to discriminate the classes. In addition, we use L2-regularization for the MLR ($C=1.0$ in the scikit-learn implementation). In contrast, the 1NN assigns to a new sample the class to which belongs its closest neighbors with respect to a similarity metric, here chosen as the Pearson correlation coefficient between the feature vectors. It thus relies on the global profile of features to cluster samples into classes.

Following standards, we use a stratified cross-validation scheme with 80\% of the samples for the train set and 20\% for the test set, where the ratio of classes are the same in both sets. We also use the subject identity as ``group information'' to avoid mixing subject data between the train and test sets. In practice, we use 50 random splits of the data and report the distribution of accuracies of the 50 splits (see the violin plots).

\renewcommand{\thefigure}{S1}
\begin{figure}
\centering
\includegraphics[width=0.98\columnwidth]{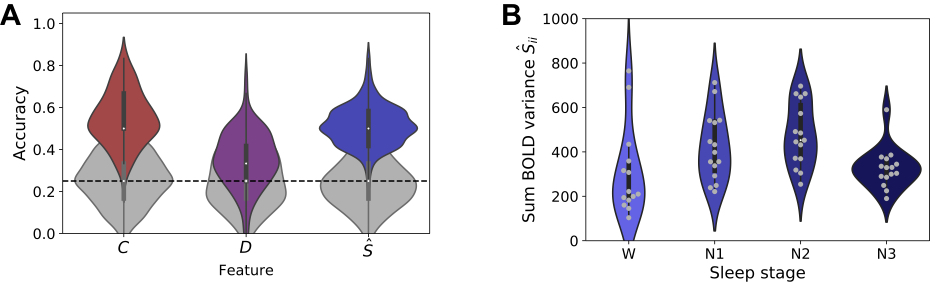}
\caption{
A) Similar plot to Fig~2C in the main text with the classification accuracy for the 1-nearest-neighbor (1NN) classifier, which relies on a similarity measure (here the Pearson correlation) between the input features to predict the class of the test sample.
The x-axis indicates the features: the model estimates $\textbf{C}$ and $\textbf{D}$, as well as the empirical FC denoted by $\hat{\textbf{S}}$.
B) Similar plot to Fig~2B in the main text but for the model input variance summed over all ROIs.}
\label{figs2_classif}
\end{figure}

Fig~\ref{figs2_classif} shows that the decoding of the sleep states by the 1NN classifier, which assigns to a new sample the class to which belongs its closest neighbors with respect to a similarity metric, here chosen as the Pearson correlation coefficient between the feature vectors. It thus relies on the global profile of features to cluster samples into classes.

\subsection{Model fitting and goodness of fit across sleep
stages.}

The model fit was quantified using two measures. The model error, defined by matrix distance, and the Pearson correlation between vectorized FC matrices (model versus data). As shown in Fig~\ref{figs1_fit}A-B, all sleep states have Pearson correlation above 0.6. Note that the changes in the goodness of fit of the model, as measured by the Pearson correlation across sleep stages in Fig~\ref{figs1_fit}B, are likely due to the stronger similarity between the corresponding empirical FC matrices in the deep sleep stages than in light sleep stages. In other words, we do not expect this trend to indicate a much better fit of the model for N3 than for W (which might have implications for the calculation of entropy production).

\renewcommand{\thefigure}{S2}
\begin{figure}
\centering
\includegraphics[width=0.98\columnwidth]{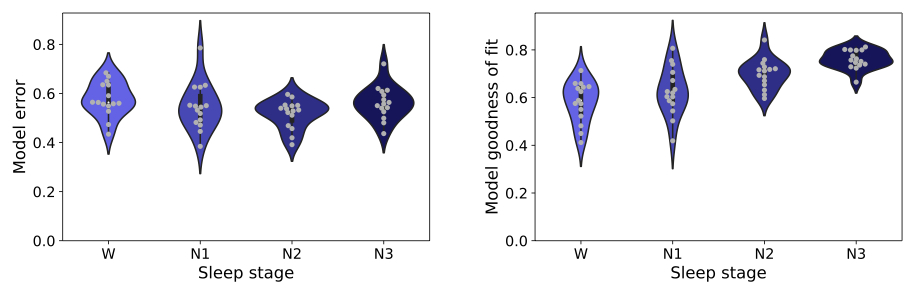}
\caption{
A: Model error for each sleep stage (x-axis) across the 15 subjects. The sleep stages are represented by the blue contrasts, from light for wake (W) to dark for the deepest sleep (N3).
B: Goodness of fit as measured by the Pearson correlation between the vectorized model and empirical FC matrices: $S(0)$ with $\hat{S}(0)$, and $S(1)$ with $\hat{S}(1)$. The average of the two Pearson correlation values is reported. C: Isomap decomposition of the empirical zero lag functional connectivity matrices FC0 (or $\hat{S}(0)$) in two dimensions to illustrate their stronger clustering in deep sleep stages N3 (squares) and N2 (triangles), compared to N1 (dots) and W (crosses). D: Pearson similarity across $\hat{S}(0)$ matrices in each sleep stage (x-axis) for all pairs of subjects}
\label{figs1_fit}
\end{figure}

\subsection{Complementary analysis}

The comparison of the nodal irreversibility across sleep states shows an overall decrease for all ROIs from W to N3 (Fig~\ref{figs3_comp_rois}A), with a pronounced decrease from W to N1 (Fig~\ref{figs3_comp_rois}B) and to a lesser extent from N2 to N3 (Fig~\ref{figs3_comp_rois}D), although N1 and N2 are rather similar (Fig~\ref{figs3_comp_rois}C). Importantly, we can see heterogeneity in the reduction of irreversibility across the ROIs in the transition to deep sleep.

\renewcommand{\thefigure}{S3}
\begin{figure}
\centering
\includegraphics[width=1\columnwidth]{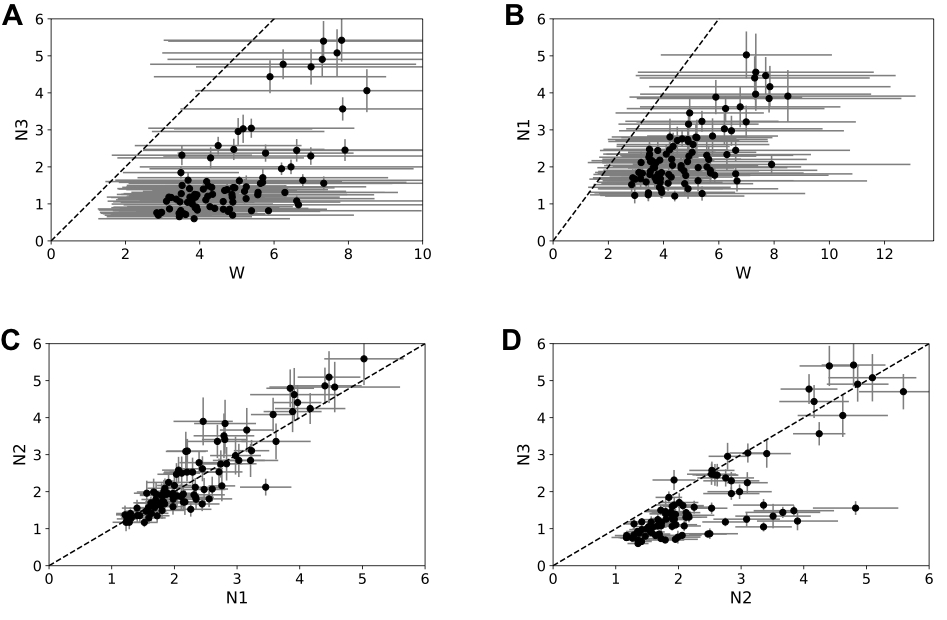}
\caption{
A) Plot of the sum of absolute values in $Q$ over each row across the W and N1 states. The error bars indicate the s.e.m. across subjects. This plot is another view of the same data in Fig~3B (main text).
B-C) Similar plots to panel A for N1 versus N2, N2 versus N3 and W versus N3.
}
\label{figs3_comp_rois}
\end{figure}

Fig~\ref{figs4_gofvsep} shows the correlation between the goodness of fit and the entropy production across all subjects for each brain state. We observe a lack of statistically significant Spearman correlations between variables except for the awake state, where the effect was close to the significance threshold. Fig~\ref{figs1_fit} shows the isomap decomposition of the empirical zero lag functional connectivity matrices FC0 in two dimensions to illustrate their stronger clustering in deep sleep stages  compared to N1 and W. We also show the Pearson similarity across  FC0 matrices in each sleep stage (to be compared with fig 3B in main text.

\renewcommand{\thefigure}{S4}
\begin{figure}
\centering
\includegraphics[width=1\columnwidth]{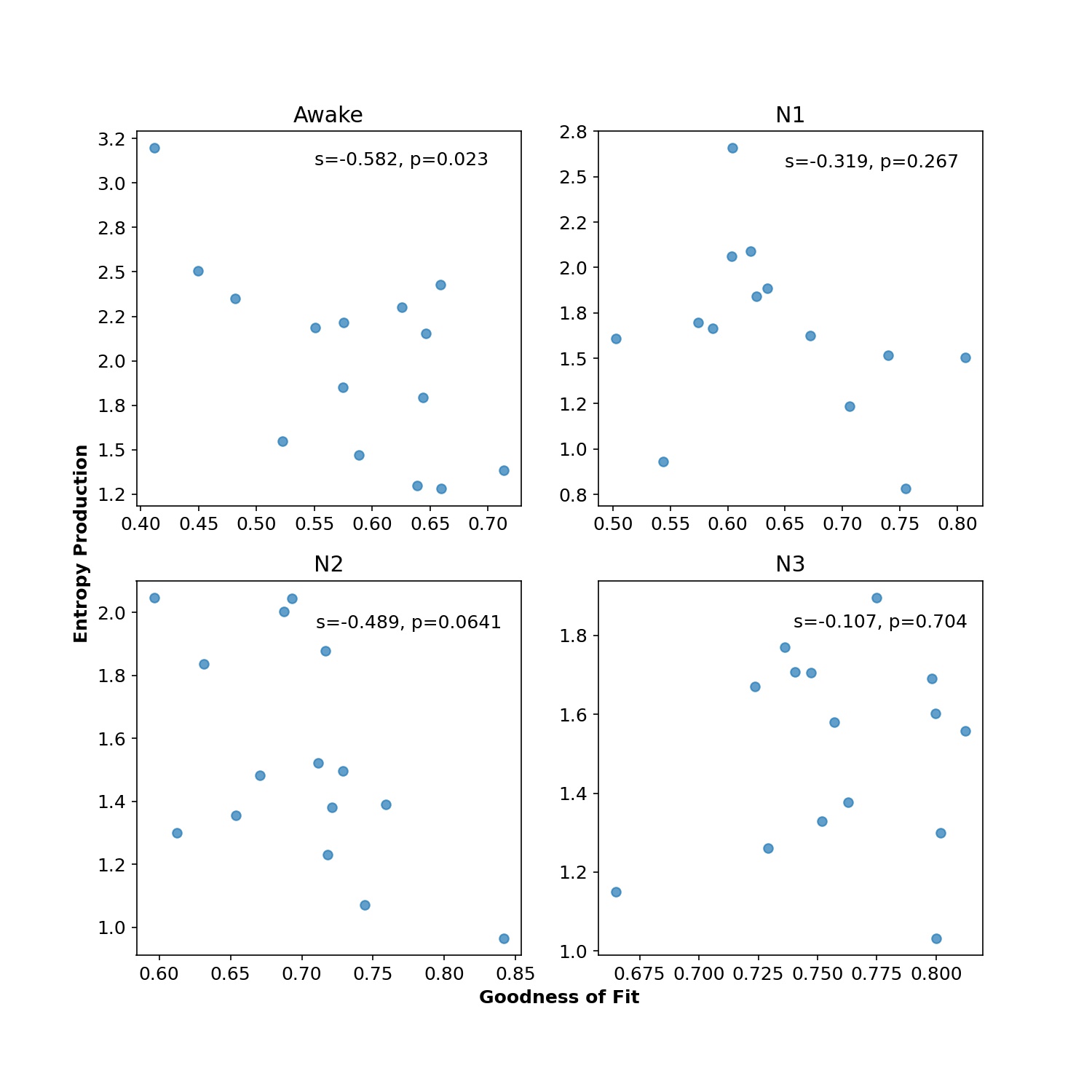}
\caption{
Correlation between the goodness of fit and the entropy production  across all subjects for each brain state. In each plot, we show the Spearman correlation, denoted $s$, with the corresponding p-value. The correlations are not statistically significant, except for the Awake state. }
\label{figs4_gofvsep}
\end{figure}

\subsection{Resting-State fMRI signals}

\subsubsection*{Participants}

A total of 63 healthy subjects ($36$ females, mean $\pm$ SD, \rev{$23.4\pm 3.3$ years}) were selected from a data set previously described in a sleep-related study by Tagliazucchi and Laufs \citep{Tagliazucchi2014}. Participants entered the scanner at 7 PM and were asked to relax, close their eyes, and not fight the sleep onset.  A total of 52 minutes of resting state activity were measured with a simultaneous combination of EEG and fMRI. According to the rules of the American Academy of Sleep Medicine 
, the polysomnography signals (including the scalp potentials measured with EEG) determine the classification of data into four stages (wakefulness, N1, N2, and N3 sleep). 
We selected 15 subjects with contiguous resting-state time series of at least 200 volumes to perform our analysis. The local ethics committee approves the experimental protocol (Goethe-Universität Frankfurt, Germany, protocol number: 305/07), and written informed consent was asked to all participants before the experiment. The study was conducted according to the Helsinki Declaration on ethical research.

\subsubsection*{MRI data acquisition}
MRI images were acquired on a 3-T Siemens Trio scanner (Erlangen, Germany) and fMRI acquisition parameters were 1505 volumes of T2-weighted echo planar images, TR/TE = 2080 ms$/30$ ms, matrix $64 \times 64$, voxel size $3$×$3$×$3$ mm$^3$, distance factor $50\%$; FOV $192$ mm$^2$. An optimized polysomnographic setting was employed (chin and tibial EMG, ECG, EOG recorded bipolarly [sampling rate $5$ kHz, low pass filter $1$ kHz] with $30$ EEG channels recorded with FCz as the reference [sampling rate 5 kHz, low pass filter 250 Hz]. Pulse oximetry and respiration were recorded via sensors from the Trio [sampling rate $50$ Hz]) and MR scanner-compatible devices (BrainAmp MR+, BrainAmpExG; Brain Products, Gilching, Germany), facilitating sleep scoring during fMRI acquisition.

\subsubsection*{Brain parcellation AAL 90 to extract BOLD time series and filtering} 
To extract the time series of BOLD signals from each participant in a coarse parcellation, we used the AAL90 parcellation with 90 brain areas anatomically defined.
BOLD signals (empirical or simulated) were filtered with a Butterworth (order 2) band-pass filter in the 0.01-0.1 \rev{Hz} frequency range.



%

\end{document}